\documentclass[prl,twocolumn,showpacs,preprintnumbers,nofootinbib,amsmath,amssymb,nobalancelastpage]{revtex4}
\usepackage{graphicx}
\usepackage{bm}
\usepackage{dcolumn}
\usepackage{amsmath}
\usepackage{amssymb}
\usepackage{color}

\usepackage{aas_macros}

\newcommand{\cm}{{\mathrm{cm}}}
\newcommand{\second}{{\mathrm{s}}}
\newcommand{\GeV}{\mathrm{GeV}}

\newcommand{\ochih}{\Omega_\chi h^2}
\newcommand{\sA}{\sigma_A}
\newcommand{\sigv}{\langle \sA v \rangle}
\newcommand{\mass}{M_\chi}
\newcommand{\phipp}{\Phi_{\mathrm{PP}}}
\newcommand{\degr}{^\circ}
\newcommand{\Aeff}{A_\mathrm{eff}}
\newcommand{\Tobs}{T_\mathrm{obs}}
\newcommand{\prob}{\mathrm{P}}
\newcommand{\Nvec}{{{\bf N}}}
\newcommand{\ethreshold}{E_{\mathrm{th}}}

\begin{document}

\title{Exclusion of canonical WIMPs by the joint analysis of Milky Way dwarfs with data from the Fermi Gamma-ray Space Telescope}
\author{Alex Geringer-Sameth}
\email{alex_geringer-sameth@brown.edu}
\author{Savvas M. Koushiappas}
\email{koushiappas@brown.edu}
\affiliation{Department of Physics, Brown University, 182 Hope St., Providence, RI 02912}

\date{\today}

\begin{abstract}
Dwarf spheroidal galaxies are known to be excellent targets for the detection of annihilating dark matter. 
We present new limits on the annihilation cross section of Weakly Interacting Massive Particles (WIMPs) based on the joint analysis of seven Milky Way dwarfs using a frequentist Neyman construction and Pass 7 data from the Fermi Gamma-ray Space Telescope.
We exclude generic WIMP candidates annihilating into $b\bar{b}$ with mass less than 40 GeV that reproduce the observed relic abundance. To within 95\% systematic errors on the dark matter distribution within the dwarfs, the mass lower limit can be as low as 19 GeV or as high as 240 GeV. For annihilation into $\tau^+\tau^-$ these limits become 19 GeV, 13 GeV, and 80 GeV respectively.

\end{abstract}

\pacs{95.35.+d, 98.80.-k, 95.55.Ka, 07.85.-m}

\maketitle

Weakly Interacting Massive Particles (WIMPs) have long been considered well-motivated and generic candidates for dark matter \cite{Zel1,Zel2,1977PhRvL..39..165L,1978ApJ...223.1015G,1978AJ.....83.1050S,1984NuPhB.238..453E}. By virtue of  weak interactions with standard model particles, WIMPs in thermal equilibrium in the early universe ``freeze out'' by the same mechanism which explains the observed abundance of light nuclei. The present-day abundance of WIMPs is governed by their annihilation cross section into standard model particles.

Due to the form of their weak-scale cross section, WIMPs have a dark matter density  $\ochih \simeq 3 \times 10^{-27} \, \cm^3 \, \second^{-1} / \sigv$, roughly irrespective of the particle mass \citep{1996PhR...267..195J}.
For the measured  $\ochih \simeq 0.1$ \cite{2011ApJS..192...18K}, the velocity-averaged annihilation cross section is  $\sigv \sim 3\times 10^{-26} \, \cm^3\,\second^{-1}$. Because a smaller cross section overproduces the observed density, this value should be seen as a relatively strong lower bound on $\sigv$ in the canonical thermal WIMP scenario.
If observations can lower the upper limit on $\sigv$ below this level, they will present a serious challenge to the conventional WIMP hypothesis (see e.g., \cite{2007PhRvL..99w1301B,2008PhRvD..78f3542M,2009JCAP...10..009C,2010arXiv1011.5090A,2010JCAP...11..041A,2010JCAP...01..031S,2010PhRvD..82l3503E,2011arXiv1110.6151A}).

It is well known that Milky Way dwarf galaxies are excellent targets to search for dark matter annihilation signatures: they are dark matter dominated objects with no astrophysical backgrounds (no hot gas). Measurements of the velocity dispersion of stars in these systems allows the reconstruction of the potential well and thus the density profile of the dark matter distribution \citep{2007PhRvD..75h3526S,2008ApJ...678..614S,2009JCAP...06..014M}.

In order to place constraints on the annihilation cross section, we must quantify how the value of $\sigv$ influences the number of $\gamma$-ray events detected with the Large Area Telescope (LAT) onboard the Fermi Gamma-ray Space Telescope (Fermi).  There are two sources of detected photon events: those arising from dark matter matter annihilation (signal), and those produced by any other processes (background). 

In the canonical picture, dark matter annihilates and gives rise to a $\gamma$-ray flux which factors into  two independent terms: one describing the dark matter particle physics and one involving the astrophysical properties of the dwarf galaxy. 
The expected number of signal events is
\begin{equation}
\mu(\phipp) \equiv (\Aeff\Tobs) \times \phipp \times J,
\label{eq:mu}
\end{equation}
where $\Aeff$ is the effective area of the detector and $\Tobs$ is the observation time. The product $\Aeff\Tobs$ is called the exposure. The goal is to place limits on the quantity $\phipp$ which encompasses the particle physics. For self-conjugate particles it is defined as
\begin{equation*}
\phipp  \equiv   \frac{\sigv}{8\pi \mass^2} \int\limits_{\ethreshold}^{\mass}  \sum_f B_f \frac{dN_f}{dE} \, dE,
\end{equation*}
where $\mass$ is the mass of the dark matter particle and $\sigv$ is its {\em total} velocity-averaged cross section for annihilation into standard model particles. The index $f$ labels the possible annihilation channels and $B_f$ is the branching ratio for each. For any channel, $dN_f/dE$ is the final $\gamma$-ray spectrum. This quantity is integrated from a threshold energy $\ethreshold$ to the mass of the dark matter particle.

The quantity $J$ contains information about the distribution of dark matter and is defined by
\begin{equation*}
J \equiv \int\limits_{\Delta \Omega(\psi)} \int\limits_{\ell} [\rho(\ell,\psi)]^2 \,d\ell \,\,d \Omega(\psi).
\end{equation*}
Here, the square of the dark matter density  is integrated along a line of sight in a direction $\psi$, and over solid angle $\Delta \Omega$. 

Typically, the background is derived through detailed modeling of  possible contributions \cite{2010ApJ...720..435A}. This was the approach taken in the Fermi Collaboration  analysis  \citep{2010ApJ...712..147A,2011arXiv1102.5701L,Garde:2011,2011arXiv1108.3546T}. In this work we eschew such detailed modeling of the origin and spectral properties of the $\gamma$-ray background, and instead use the photon events in the region near each dwarf  to empirically derive the background from all unresolved sources.

The fundamental assumption of our strategy is this: {\em whatever the processes are which give rise to the photon events nearby each dwarf, these same processes are also at work in the direction of the dwarf.} That is, the probability that background processes produce photons at the location of a dwarf can be determined by the empirical probability distribution found by sampling the observed counts in the surrounding region. The region surrounding each dwarf is a ``sideband'' used to determine the background. This approach requires zero free parameters and the entire analysis depends only on the value of $\phipp$.

We use Bootes I, Draco, Fornax, Sculptor, Sextans, Ursa Minor, and Segue 1 because none are in a crowded field or near known $\gamma$-ray sources.
We utilize the updated values of $J$ presented in \citep{2011arXiv1108.3546T}. The $J$ values are derived based on  modeling  the velocity dispersion profiles of stars in each dwarf \citep{2007PhRvD..75h3526S,2008ApJ...678..614S,2009JCAP...06..014M}.

We define a Region of Interest (ROI) to be a region of the sky with a radius of 0.5$\degr$ containing all {\tt Pass 7} photons of {\tt evclass=2} available publicly on the Fermi Science Support Center (FSSC) \cite{FSSC}, in the Mission Elapsed Time interval of [239557417-334619159] seconds (August 4, 2008 15:43:36 UTC  to August 9, 2011 21:45:57 UTC), and with energies [1-100] GeV (at these energies, the point spread function (PSF) is always less than 1$\degr$).  For each ROI, we use the publicly available version {\tt v9r23p1} of the {\tt Fermi Science Tools} to extract photons (with {\tt zmax=100}), select good time intervals (with all standard recommendations as stated on the FSSC),  and compute the exposure $(\Aeff \Tobs)$, which also takes into account the shape of the PSF within the ROI using the Instrument Response Function {\tt P7SOURCE\_V6}. Because the  PSF is energy dependent, the exposure must be averaged with the  annihilation energy spectrum. 
For a range of  power-law indices of the  spectrum the exposure within an ROI changes by at most 5\%, making this a negligible effect in the cross section limits.

We identify and mask all sources present within 10$\degr$ of each dwarf using the 2nd Fermi Source Catalogue \cite{Fermi2c} (with a masking size of 0.8$\degr$). We calculate the probability of observing background events at the location of the dwarf by sampling $10^5$ ROIs which are randomly selected within a distance of $10\degr$ from each dwarf, and counting the events in each. A window is rejected if it overlaps with a masked location or with the boundary. There are approximately $(10 / 0.5)^2 = 400$ independent ROIs for each dwarf. 
The background probability mass function (PMF) is  given by the fraction of ROIs that contained a given number of counts (the PMF is not sensitive to increasing the mask size to 2$\degr$). This PMF is taken to be the probability distribution governing the number of background photons which contribute to the central ROI. 

The accuracy of this strategy requires   the total exposure  not vary within a $10\degr$ radius around each dwarf and we find that it varies by at most $\sim 5\%$. 
If a $\gamma$-ray source is  close to a dwarf it may contribute photons to the central ROI. These source photons are not accounted for in the empirical background PMF. Therefore, such photons are considered more likely to be from dark matter annihilation and will weaken the derived limit. In this sense, our analysis is conservative. 
The PMFs are well fit by Poisson distributions and do not contain  features that would be expected from source contamination \footnote{See appendix for the empirically derived PMFs, weights, and counts in the central ROI for the dwarfs used in this analysis.}.

In statistical inference one wants to generate confidence intervals for a model parameter $\mu$ based on  observed data $x$. In a frequentist analysis the main task is to decide on an {\em algorithm} which constructs a region in $\mu$-space for any value of $x$. This region is said to be an $\alpha$-confidence interval if the algorithm has ``coverage'' $\alpha$ (see e.g. \cite{Kendall5th,1998PhRvD..57.3873F}).

One way to construct and visualize confidence intervals is by using the Neyman construction \citep{1998PhRvD..57.3873F, 1937RSPTA.236..333N}. The ingredients needed are the parameter space of possible $\mu$ values, a space of possible measurements $x$, and a likelihood function $\prob(x|\mu)$, which gives the probability of observing $x$ if $\mu$ were the true value of the parameter ($\mu$ and $x$ can both live in any number of dimensions). For each possible value of the parameter $\mu$ one selects a region $D(\mu)$ of the measurement space such that $\int_{D(\mu)} \prob(x|\mu) = \alpha$ (i.e. the probability of measuring $x$ to be in $D(\mu)$ is $\alpha$ if the true value of the parameter were $\mu$). These regions are called confidence belts. For an actual measurement $x^*$, these pre-selected belts can be used to generate an $\alpha$-confidence region for $\mu$. This region is simply the  collection of all the $\mu$ values whose belt $D(\mu)$ contains $x^*$. This algorithm for constructing a region in $\mu$-space out of a measured value $x^*$ provides the proper coverage: whatever the true value $\mu_t$ is, there is an $\alpha$ chance that $x^*$ will lie in $D(\mu_t)$ (by construction) and therefore an $\alpha$ chance that the resulting confidence interval will contain $\mu_t$.

In this analysis, the observations consist of the number of counts $N_i$ from the central ROI containing each dwarf $(i=1,\dots,7)$. These can be considered the components of a vector $\Nvec$ living in a 7-dimensional integer lattice. To apply the Neyman construction we must choose a confidence belts in this 7-dimensional ``$N$-space'' for every possible value of $\phipp$, such that the probability that $\Nvec$ is measured to be in this belt is $\alpha$.

There is complete freedom in the choice of belts (provided they have coverage $\alpha$). Nevertheless, it is vital that the shape of the belts for each $\phipp$ not be based on the measured data. This offense is known as ``flip-flopping" \cite{1998PhRvD..57.3873F}. It may result in confidence levels having lower coverage than stated. Here, the confidence belts are constructed {\it without prior knowledge} of the number of counts within the central ROI around each dwarf. Under the assumption that the empirically derived background PMFs, exposures, and J values are correct, the belts have the proper coverage.

\begin{figure}
\includegraphics[scale=0.92]{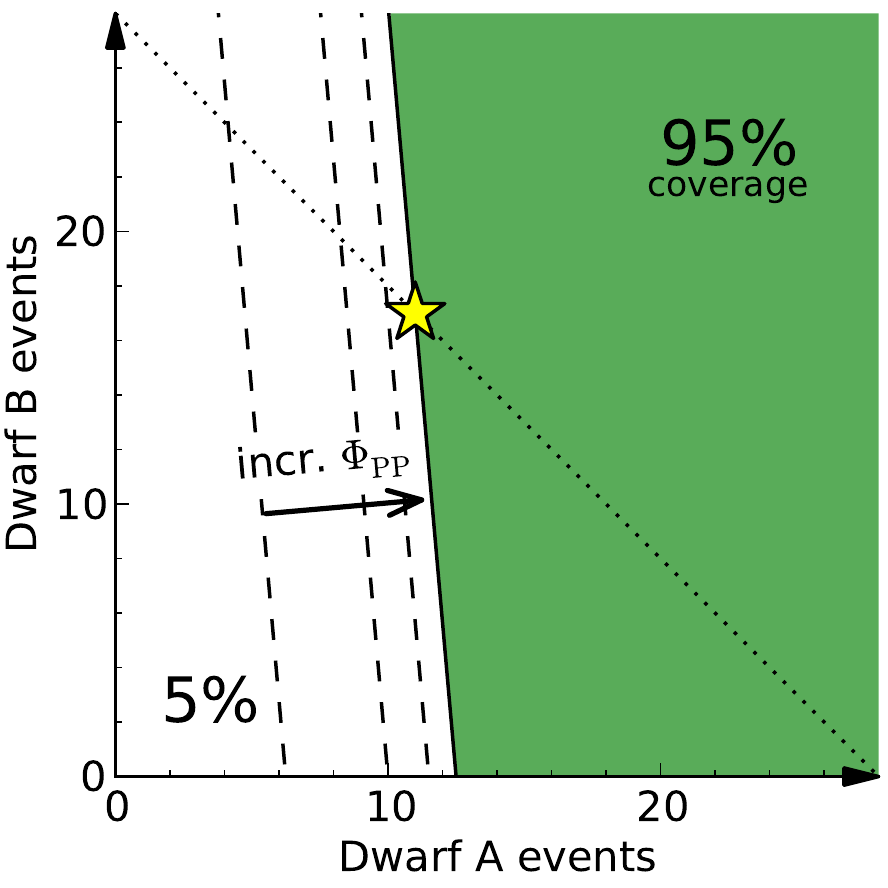}
\caption{\label{fig:Nspace} Illustration of the Neyman confidence belt construction used to generate upper limits on $\phipp$. Each axis represents the number of events that could be observed from a given dwarf  (here, Dwarf A has a larger $J$ value than Dwarf B does). The shaded area, bordered by the solid line, represents the confidence belt for a particular value of $\phipp$. The dashed lines are the borders of the confidence belts for different values of $\phipp$, with $\phipp$ increasing from left to right. The borders are chosen to be normal to a vector of ``sensitivities'', which weights each dwarf according to the relative strength of its dark matter signal. Once a measurement is made (shown by the star) the confidence interval for $\phipp$ contains all values of $\phipp$ whose confidence belt contains the measured point. The dotted line shows the border for an alternative construction of the confidence belts which gives equal weight to each dwarf.}
\end{figure}

In order to derive an upper limit on $\phipp$, the $N$-space should be divided into two simple parts and the belt $D(\phipp)$ should consist of the ``large'' $\Nvec$ values (i.e. the region containing $N_i = \infty$). This is illustrated in Fig.~\ref{fig:Nspace} for an example joint analysis of two dwarfs. 
The simplest choice for the confidence belt boundaries are planes with normal vectors parallel to $(1,\dots,1)$, represented in Fig.~\ref{fig:Nspace} by the dotted line. A measured set of $N_i$ is in such a confidence belt if the sum of the $N_i$ is greater than some value. This is equivalent to ``stacking'' the events from each dwarf and then analyzing this single image. However, because the dwarfs are treated equally, photons from a dwarf with a small $J$ value are considered as likely to have come from dark matter as are photons from a dwarf with large $J$. This is  an inefficient choice for the confidence belts. Naively, one extra photon from Draco ($J \propto 0.63$) should raise the upper limit  more than an extra photon from Bootes I ($J \propto 0.05$) because, a priori, a given photon from Bootes I is much more likely to be from background than a photon from Draco.

To overcome this obstacle we take advantage of the recent idea by \citet{2009CQGra..26x5007S} to use planes at angles other than $45\degr$ as boundaries of the confidence belts. Sutton suggests letting the normal vector to the planes be equal to a vector representing the ``sensitivity'' of each observation. We take the sensitivity (or weight) of each dwarf observation to be proportional to the ratio of the expected dark matter flux $(\Aeff\Tobs\,J)$ to the mean expected empirical background flux.
In contrast, giving every dwarf the same weight can weaken the   limits by as much as 25\%.

The number of photons received in the central ROI containing each dwarf is the sum of the number of photons  from dark matter annihilation and the number produced by all background processes. The number of signal photons is governed by a Poisson distribution with mean $\mu(\phipp)$ (Eq.~\ref{eq:mu}).  The number of background photons  is described by the empirical background PMF. Therefore, the total number of photons detected is distributed according to the convolution of these two probability distributions.  The counts found for each dwarf  are independent variables and so the joint probability of measuring $\Nvec$ is given by the product of the individual PMFs.


\begin{figure}
\includegraphics[scale=0.42]{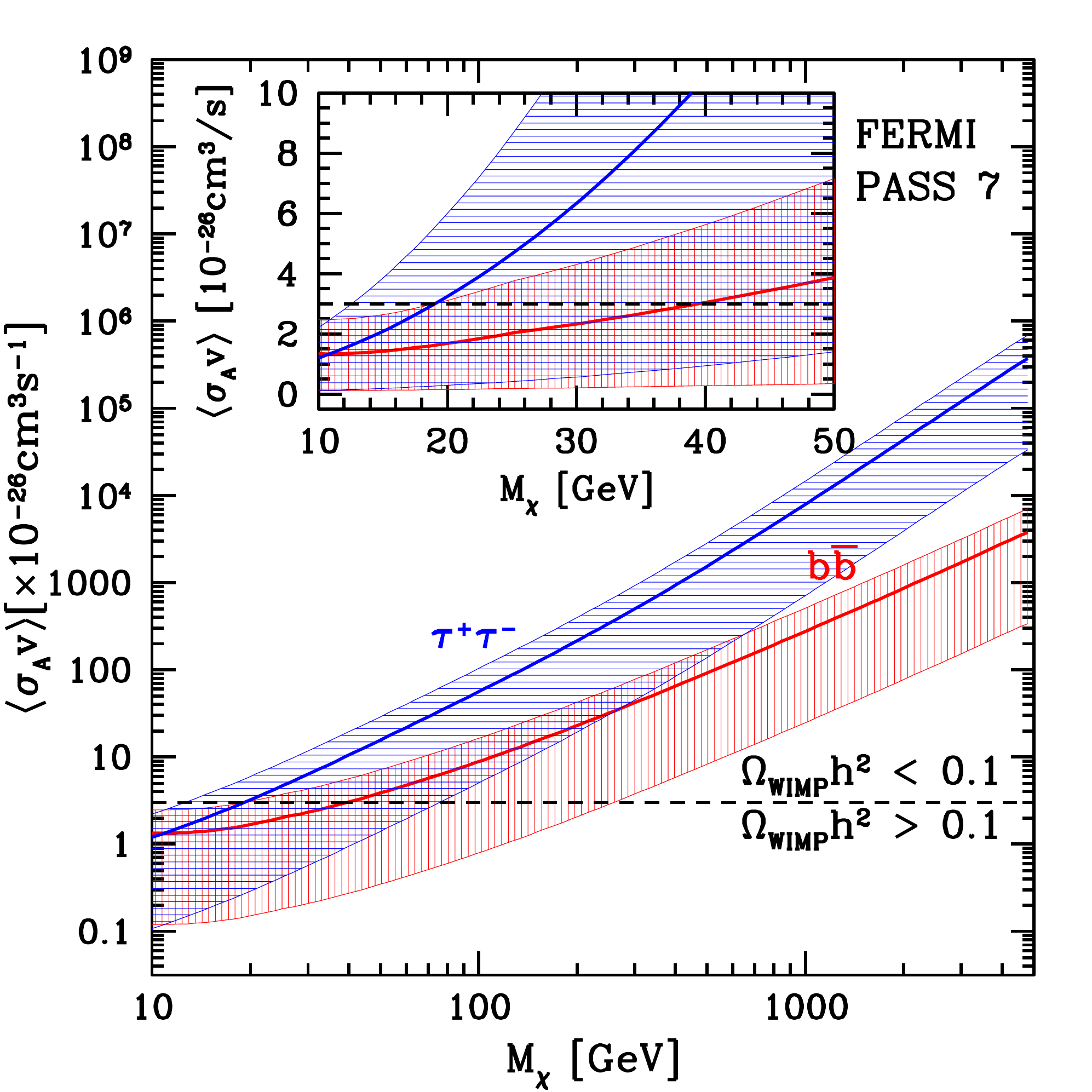}
\caption{\label{fig:sigvlim} Derived 95\% upper limit on $\sigv$ as a function of mass for dark matter annihilation into $b{\bar b}$ and $\tau^+\tau^-$. The shaded area reflects the 95-percentile of the systematic uncertainty in the dark matter distribution of the dwarfs. The canonical annihilation cross section for a thermal WIMP making up the total observed dark matter abundance is shown by the dashed line. The inset figure shows detail for lower masses.}
\end{figure}

Using this statistical framework we derive a 95\% upper bound of $\phipp = 5.0 ^{+4.3}_{-4.5}\times10^{-30}\,\cm^3\, \second^{-1}\, \GeV^{-2}$. In order to translate the bound on $\phipp$ into a bound on $\sigv$ as a function of $\mass$ we need to assume a specific annihilation channel and its spectrum $dN/dE$. It is generally assumed that a WIMP annihilates primarily into hadrons (e.g. $b{\bar b}$) or heavy leptons (e.g. $\tau^+\tau^-$), which then decay by fairly well constrained channels into $\gamma$-rays. We compute $dN/dE$ for these channels using DarkSUSY \cite{2004JCAP...07..008G,darksusy}.

Figure~\ref{fig:sigvlim} shows the derived 95\% upper bound on $\sigv$ as a function of WIMP mass. For annihilation into $b \bar{b}$ ($\tau^+\tau^-$) WIMP masses less than 40 GeV (19 GeV) are excluded using the central $J$ values\footnote{See appendix for limits on additional annihilation channels.}.  
The dominant source of systematic uncertainty comes from the poorly constrained $J$ for each dwarf and is shown by the shaded regions in Fig.~\ref{fig:sigvlim}. The $\phipp$ limit is recalculated for each dwarf as its $J$ varies between its upper and lower 95\% error bar given in \cite{2011arXiv1108.3546T}. The results for each dwarf are then added in quadrature (this procedure gives a nearly identical region as that derived by scanning over the log-normal priors on $J$ for each dwarf \cite{2007PhRvD..75h3526S,2008ApJ...678..614S,2011arXiv1108.3546T}). 

If we knew the exact $J$ value of each dwarf, the width of the shaded regions in Fig.~\ref{fig:sigvlim} would shrink to zero. 
However, due to the uncertainties in $J$, we have no knowledge of where this upper limit lies within the shaded region. 
Presenting the limit in this fashion  separates the inherent statistical uncertainties (Poisson-distributed photon counts) from the systematic errors in the $J$'s, which in principle could be known exactly (each dwarf has ``a" dark matter distribution).   
At the present time there is no  consensus on the dark matter distribution within Milky Way dwarfs.
The systematic error bands should be thought of as an exploration of possible models for the dark matter distribution (for an alternative analysis of $J$ values see \citet{2011arXiv1104.0412C}). 
Nevertheless, for any model (set of $J$ values) the construction presented here gives a rigorous 95\% upper limit on $\phipp$. 

For the most (least) conservative model the lower limit on the mass is 19 GeV (240 GeV) for $b \bar{b}$, while for $\tau^+\tau^-$ these limits are 13 GeV (80 GeV). 
Segue 1 is responsible for most of the uncertainty in the limit due to its high weight and uncertain dark matter content. However, if Segue 1 has a low $J$ value, the statistical construction downgrades its weight relative to other dwarfs such as Draco and Ursa Minor.

The strength of the analysis relies on the validity of the assumption that the background at the location of each dwarf is adequately described by the empirical PMF. In general, if the assumed background PMF is skewed toward higher numbers of counts the upper limit on $\phipp$ becomes stronger. This is because more of the observed counts can be attributed to background and therefore fewer to dark matter annihilation. We can quantify the effect of an error in the empirical PMF by considering the radical case where we are certain there is no background at all. This is a false assumption, but is one which will produce the most conservative limit on $\phipp$. If we force the background PMFs to be equal to 1 when the number of counts is 0 and 0 otherwise, the 95\% limit on $\phipp$ increases by a factor of 4.4 over the actual limit. This represents the case where every photon received from the dwarf is believed to be due to dark matter annihilation. We interpret this as a test of the robustness of the method not as any sort of actual confidence limit.
We can also test our conclusions against less violent changes to the background PMF. For each dwarf we replace the background PMF with a Poisson distribution having the same mean, and find that the limit on $\phipp$ decreases by 7\%.

What is the significance of this new bound on $\sigv$? It signals, perhaps, that we are imminently approaching an epoch of discovery. Three decades of experimental design have given rise to many detectors sensitive enough to probe a very generic class of dark matter candidates. The prime motivation for WIMP dark matter is the coincidence that a weak-scale annihilation cross section naturally reproduces the observed relic abundance. Unlike the scattering cross section probed in direct detection experiments, cosmology gives a lower limit for the annihilation cross section. The parameter space in which a WIMP can hide is therefore bounded at both ends. This work, together with the Fermi-LAT collaboration result \cite{2011arXiv1102.5701L,2011arXiv1108.3546T,Garde:2011}, pushes the contact point between the upper and lower bounds on $\sigv$ to increasing WIMP masses, suggesting that  observations have become powerful enough to either discover or rule out the best-motivated and most sought-after dark matter candidates.

\begin{acknowledgments}
We acknowledge useful conversations with John Beacom, Richard Gaitskell, Elizabeth Hays, Andrew Hearin, Robert Lanou, Julie McEnery, Tim Tait and Andrew Zentner. AGS and SMK are supported by NSF grant PHY-0969853 and by Brown University.
\end{acknowledgments}

\bibliography{manuscript}

\clearpage

\begin{figure}
\includegraphics[scale=0.95]{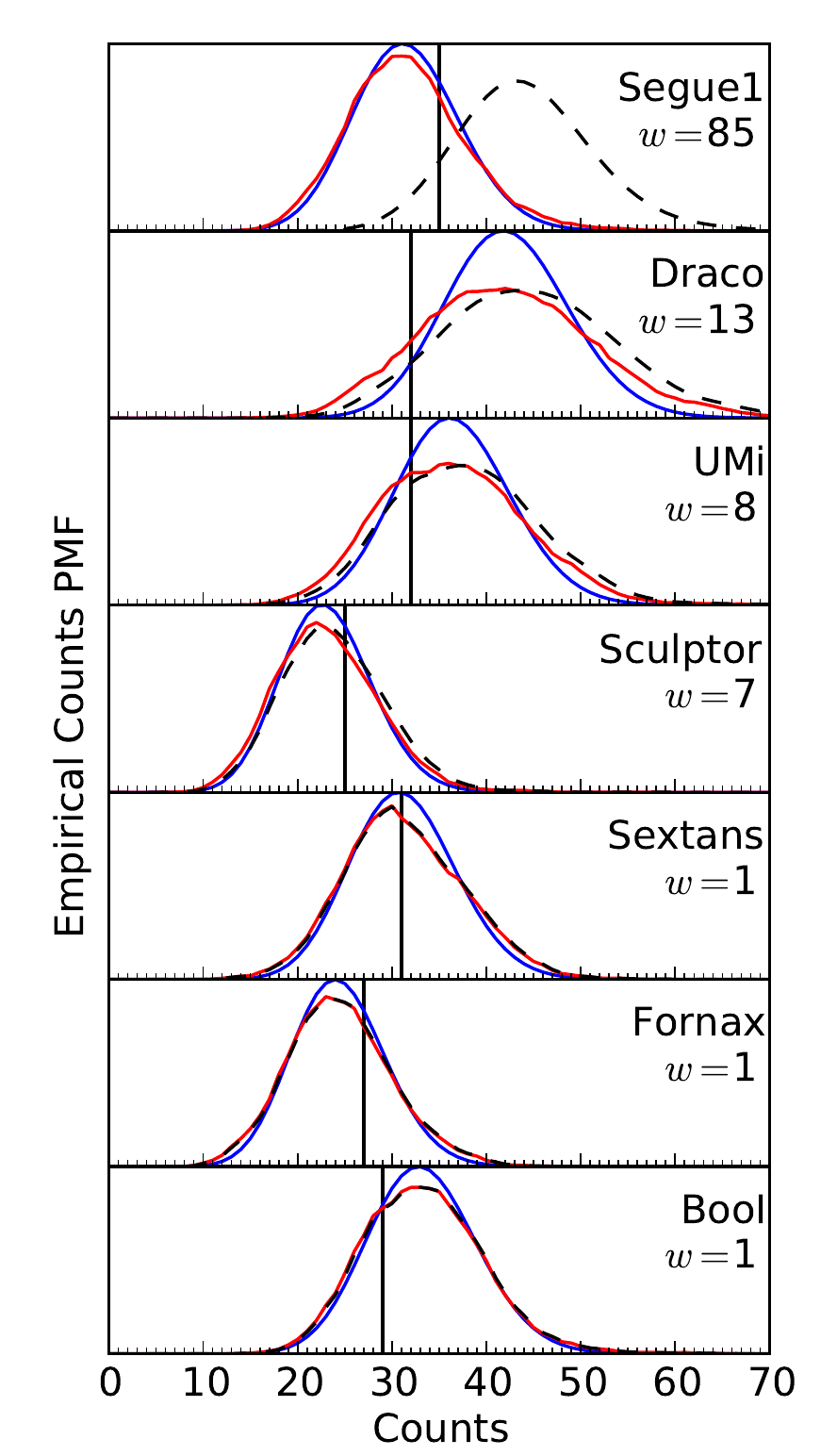}
\caption{This figure illustrates the ingredients and data required to derive upper limits on the dark matter annihilation cross section. Each plot corresponds to a different dwarf galaxy. Sampling the counts in 0.5$^\circ$ regions surrounding each dwarf results in an empirical background probability mass function (PMF) shown in red. The blue curves are Poisson distributions having the same mean as the empirical background PMFs. The vertical line represents the number of counts observed in the ROI centered on the dwarf's location. The dashed curve is the convolution of the background PMF with the Poisson distribution representing the contribution from dark matter annihilation when $\phipp = 5.0 \times10^{-30}\,\cm^3\, \second^{-1}\, \GeV^{-2}$ (the 95\% upper limit on $\phipp$). This convolution is the probability distribution of the sum of signal and background. The label $w$ is the weight given to each dwarf in our construction of Neyman confidence belts. It is given by the ratio of the strength of the expected dark matter signal to the mean expected background.}
\end{figure}

\begin{figure}
\includegraphics[scale=.4]{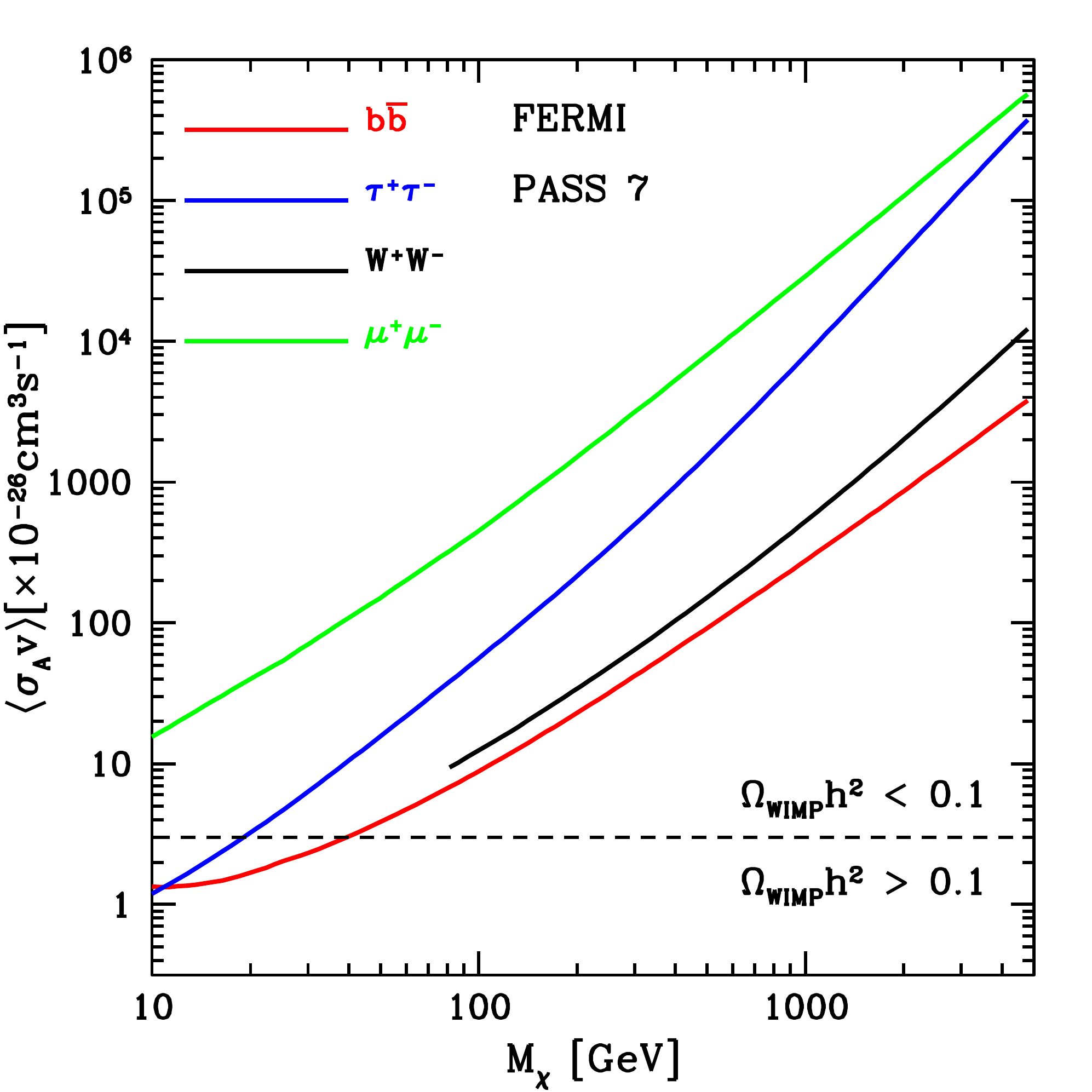}
\caption{Derived limits on $\sigv$ as a function of mass for dark matter annihilation into $W^+W^-$, a heavy quark final state ($b{\bar b}$), and heavy lepton final states $\tau^+\tau^-$ and $\mu^+ \mu^-$.  Limits on other heavy quark final states are similar to the $b{\bar b}$ channel. Annihilation spectra are derived using \cite{2004JCAP...07..008G,darksusy}. All limits are 95\% upper limits based on the central $J$ values for the dwarfs. }
\end{figure}

\end{document}